\documentclass[12pt]{article}
\usepackage{epsfig,econf}
\index{<item name>}
\def\Title#1{\begin{center} {\Large {\bf #1} } \end{center}}
\begin{document}
\Title{A Time Dependence of QCD}
\bigskip\bigskip
\begin{raggedright}  
{\it Harald Fritzsch\index{Reggiano, D.}\\
Department f\"ur Physik\\
Ludwig--Maximilians--Universit\"at\\
D-80333 M\"unchen}
\bigskip\bigskip
\end{raggedright}
\begin{abstract}
From astrophysics there are indications that the finestructure constant
$ \alpha$ has changed during the past 10 billion years. Within grand
unification one can deduce that also the QCD scale has changed. Tests
for a time variation of this scale  are described. The result of the new
experiment in Munich is discussed.
\end{abstract}
The theory of QCD is very remarkable. It is a theory of very few parameters,
i. e. only $\Lambda_c$ and the quark masses. The latter are related
to inputs by the flavor interactions and have nothing to do with the strong
interactions. The parameter $\Lambda_c$ just sets the scale of the strong
interactions and is not a real parameter for the strong interaction itself.
Thus the QCD--theory, proposed by Gell--Mann and myself in 1972
\cite{FritGell}, is indeed an exceptional theory, describing lots of
complexities in terms of very few parameters, which, as discussed below,
might even depend on time.

Usually in physics, especially in particle physics, we deal with the local
laws of nature, say the field equations of QCD or the Maxwell equations.
But when it comes
to the fundamental constants, like the finestructure constant $\alpha $,
we must keep in mind that also questions about the boundary conditions of the
universe come up. We do not know, where these constants, like $\alpha $ or
$\alpha_s $ or the lepton and quark masses, come from, but it could well
be that at least a few of them are products of the Big Bang. If the
Bing Bang would be repeated, these constants could easily take different
values. But in this case it is clear that the constants could
never be calculated.

So in connection to the fundamental constants the question comes up, whether
they are really cosmic accidents, or whether they are determined by the
dynamics, whether they are changing in time or in space, or whether they are
indeed calculable in a hypothetical theory going far above the present
Standard Model. Also considerations related to the Anthropic Principle
should be made. Life in our universe can exist only if the values of the
fundamental constants take on certain values. In a universe in which, for
example, the $u$--quark is heavier than the $d$-quark, the proton would
decay in a neutron, and life would not exist, at least not in a form
known to us.

Of course, today $\alpha $ is just the interaction constant, describing
e. g. electron--scattering at low energies:
\begin{equation}
\alpha^{-1} = 137.03599976 \ .
\end{equation}

But it is remarkable. Based on this number, one can calculate all effects in
QED to an accuracy of about 1 : 10.000.000, e. g. the magnetic moment of
the electron. Of course, QED is only a part of the Standard Model of today,
based on a superposition of QCD and the $SU(2) \times U(1)$ -- electroweak
theory, and $\alpha $ is just one of at least 18 parameters, entering the
Standard Model.

One of the fundamental quantities is the proton mass. I should like to stress
that the proton mass is a rather complicated object in the Standard Model.
The coupling constant of QCD follows in leading order the equation:
\begin{equation}
\alpha_s \left( Q^2 \right) = \frac{2 \pi}{b_0} \, \, \, {\rm ln} \, \, 
\left( \frac{Q}{\Lambda} \right), b_0 = 11 - \frac{2}{3} n_f \, .
\end{equation}

Here the scale parameter $\Lambda $ enters, which has been determined to be:
\begin{equation}
\Lambda = 214^{+38}_{-35} \, \, \, {\rm MeV} \, .
\end{equation}
$\Lambda $ is a free parameter of QCD, and all numbers of QCD scale
with $\Lambda $, at least in the limit where the masses of the quarks are
set to zero. But $\Lambda $ can be expressed in terms of MeV, i. e. it
is given in reference to the electron mass, which is outside QCD. The
physical parameters like the proton mass are simply proportional to
$\Lambda $, apart from a small correction due to quark masses.
The scale of confinement of the quarks is inversely proportional
to $\Lambda $.

I should also remind you that Grand Unification imposes that the
parameters $\alpha_s$, $\alpha$ and $\alpha_w$ are not independent. They are
related to each other, and related to the unified coupling constant,
describing the interaction at the unification scale $\Lambda_{{\rm un}}$.

It is known that the group $SU(5)$ does not describe the
observations, since the three coupling constants do not converge precisely.
If supersymmetric particles are added at an energy scale of about 1 TeV,
a convergence takes place, however \cite{Amaldi}. In $SO(10)$, proposed
by P. Minkowski and me  \cite{Fritzsch} the
situation is different, since in this group the unification is a
two--step process, where another mass scale, the mass scale for the
righthanded $W$--boson, enters. If this mass scale is chosen in the
right way, the unification can be achieved without supersymmetry.

After these preparations let me come to the question of time dependence.
A group of physicists has recently
published their evidence that the finestructure constant had a different value
billions of years ago \cite{Webb}. They were investigating the light
from about 134 quasars, using the so--called ``many multiplet method''. They
were looking at the fine--structure of atomic lines, originating from
elements like Fe, Ni, Mg, Sn, Ag etc. \, .

One particular aspect is that the fine--structure is a rather complex
phenomenon, fluctuating in particular also in the sign of the effect. These
sign changes have been observed and used in fixing the experimental values
of $\alpha $. The result is:
\begin{equation}
\frac{\Delta \alpha}{\alpha} = \left( - 0.72 \pm 0.18 \right)
\cdot 10^{-5} \, .
\end{equation}

Thus $\alpha $ was slightly larger in the past. If one takes a linear
approximation and uses a cosmic lifetime of 14 billion years, the
effect is $\dot{\alpha} / \alpha  \approx 1.2 \cdot 10^{-15}$ per year.

If $\alpha $ depends on time, the question arises, how this time--variation
is generated. Since $\alpha = e^2 / \hbar c$, a time variation could come
from a time variation of $\hbar $ or $c$. Both cases are, I think, not very
likely. If $c$ depends on time, it would mean, that we have a serious
problem with relativity. If $\hbar $ would depend on time, atomic physics
runs into a problem. So I think that a time dependence of $\alpha $ simply
means that $e$ is becoming time--dependent.

Let me also mention that according to the results of 
Dyson and Damour \cite{Damour1} there is a rather strong constraint on a
time--variation of $\alpha $, derived from the investigation of the remains of
the Oklo reactor in Gabon. If no other parameters change as well, the
relative change $\left( \dot{\alpha} / \alpha \right)$ per year cannot be
more than $10^{-17}$, i. e. there is a problem with the astrophysical
measurements, unless the rate of change for $\alpha $ has become less during
the last 2 billion years. The constraint is derived by looking at the position
of a nuclear resonance in Samarium, which cannot have changed much during the
last 2 billion years. However, I tend not to take this constraint very
seriously. According to the Grand Unification $\alpha_s$ and $\Lambda $
should have changed as well, and the two effects (change of $\alpha$ and
of $\Lambda $) might partially cancel each other.

The idea of Grand Unification implies that the gauge group $SU(3)$
of the strong interactions and the gauge group $SU(2) \times U(1)$ of the
electroweak sector are subgroups of a simple group, which causes the
unification.

Both the groups $SU(5)$ and $SO(10)$ are considered in this way.
I like to emphasize that the group $SO(10)$ has the nice property that all
leptons and quarks of one generation are described by one representation,
the 16--representation. It includes a righthanded neutrino, which does not
contribute to the normal weak interaction, but it is essential for the
appearance of a mass of the neutrino, which is expected in the
$SO(10)$--Theory.

In $SU(5)$ two representations of the group are needed to describe the leptons
and quarks of one generation, a $10$-- and a $(\bar 5)$--representation.

I should also like to emphasize that the gauge couplings $\alpha_s, \alpha_w$
and $\alpha $ meet in the $SU(5)$--theory only, if one assumes that above
about 1 TeV supersymmetry is realized. In the $SO(10)$--theory this is not
needed. A convergence of the coupling constants can be achieved, since at
high energies another energy scale enters, which has to be chosen in a
suitable manner.

A change in time of $\alpha $ can be obtained in two different ways. Either
the coupling constant $\alpha_{{\rm un}}$ stays invariant or the unification
scale changes. I consider both
effects in the $SU(5)$--model with supersymmetry. In this model the relative
changes are related:
\begin{equation}
\frac{1}{\alpha} \frac{\dot{\alpha}}{\alpha} = \frac{8}{3} \frac{1}{\alpha_s}
- \frac{10}{\pi} \frac{\dot{\Lambda}_{un}}{\Lambda_{un}}
\end{equation}

One may consider the following scenarios:
\begin{enumerate}
\item[1)] $\Lambda_G$ invariant, $\alpha_u =\alpha_u (t)$. This is the
  case considered in \cite{Calmet} (see also \cite{Lang}), and one finds
  \begin{eqnarray}  \label{eq8}
  \frac{1}{\alpha} \frac{\dot{\alpha}}{\alpha}=
\frac{8}{3} \frac{1}{\alpha_s} \frac{\dot{\alpha}_s}{\alpha_s}
\end{eqnarray}
and
\begin{equation}
\frac{\dot{\Lambda}}{\Lambda} \approx 39 \cdot \frac{\dot{\alpha}}{\alpha}
\end{equation}
\item[2)] $\alpha_u$ invariant, $\Lambda_G =\Lambda_G (t)$. One finds
  \begin{eqnarray} \label{eq10}
   \frac{1}{\alpha} \frac{\dot{\alpha}}{\alpha}=
-\frac{1}{2 \pi} \left(b_2^S+\frac{5}{3} b_1^S\right)
\frac{\dot{\Lambda}_G}{\Lambda_G}, 
\end{eqnarray}
  \begin{eqnarray} \label{eq12}
  \frac{\dot{\Lambda}}{\Lambda}= \left( \frac{b_3^{S}}{b_3^{SM}}
  \frac{1}{\alpha} \frac{\dot{\alpha}}{\alpha} \right) \approx -30.8
  \frac{\dot{\alpha}}{\alpha}
    \end{eqnarray}
  \item[3)] $\alpha_u =\alpha_u (t)$ and $\Lambda_G =\Lambda_G (t)$.
    One finds
\begin{eqnarray}
    \frac{\dot{\Lambda}}{\Lambda}& \cong &
46 \frac{\dot{\alpha}}{\alpha} + 1.07 \frac{\dot{\Lambda}_G}{\Lambda_G}
\nonumber
 \end{eqnarray}
where theoretical uncertainties in the factor
$R=(\dot{\Lambda}/\Lambda)/(\dot{\alpha}/\alpha)=46$ have been
discussed in \cite{Calmet}. The actual value of this factor is
sensitive to the inclusion of the quark masses and the associated
thresholds, just like in the determination of $\Lambda$. Furthermore
higher order terms in the QCD evolution of $\alpha_s$ will play a
role. In ref. \cite{Calmet} it was estimated: $R=38\pm 6$.
\end{enumerate}
 
According to {\cite{Calmet} the relative changes of $\Lambda$ and
$\alpha$ are opposite in sign. While $\alpha$ is increasing with a
rate of $1.0 \times 10^{-15}/$yr, $\Lambda$ and the nucleon mass are
decreasing, e.g. with
a rate of $1.9 \times 10^{-14}/$yr. The magnetic moments of the
proton $\mu_p$ as well of nuclei would increase according to
\begin{eqnarray}
\frac{\dot{\mu}_p}{\mu_p} = 30.8 \frac{\dot{\alpha}}{\alpha} \approx
3.1 \times 10^{-14}/ {\mbox yr}.
\end{eqnarray}
       
The time variation of the ratio $M_p/m_e$ and $\alpha$ discussed here
are such that they could by discovered by precise measurements in
quantum optics. The wave length of the light emitted in hyperfine
transitions, e.g. the ones used in the cesium clocks being
proportional to $\alpha^4 m_e/\Lambda$ will vary in time like
\begin{eqnarray}
\frac{\dot{\lambda}_{hf} }{\lambda_{hf}} = 4 \frac{\dot \alpha}{\alpha}
-\frac{\dot \Lambda}{\Lambda}\approx 3.5 \times 10^{-14}/\mbox{yr}
\end{eqnarray}
taking $\dot{\alpha}/\alpha\approx 1.0 \times 10^{-15}/$yr. The wavelength
of the light emitted in atomic transitions varies like $\alpha^{-2}$:
\begin{eqnarray}
\frac{\dot{\lambda}_{at} }{\lambda_{at}} = -2 \frac{\dot{\alpha} }{\alpha}.
\end{eqnarray}
One has ${\dot{\lambda}_{at} }/{\lambda_{at}}\approx
-2.0\times 10^{-15}/$yr. A comparison gives:
\begin{eqnarray}
  \frac{\dot{\lambda}_{hf}/\lambda_{hf}}{\dot{\lambda}_{at}/\lambda_{at}} = 
  -\frac{ 4 \dot{\alpha}/ \alpha - \dot \Lambda /
\Lambda}{2 \dot{\alpha}/ \alpha } \approx -17.4.
\end{eqnarray}


At present the time unit second is defined as the duration of
6.192.631.770 cycles of microwave light emitted or absorbed by the
hyperfine transmission of cesium-133 atoms. If $\Lambda$ indeed
changes, as described above, it would imply that the time
flow measured by the cesium clocks does not fully correspond with the
time flow defined by atomic transitions.

Recently a high precision experiment was done at the MPQ in Munich, using the
precise cesium clock PHARAO from Paris \cite{haensch}.

In this experiment the drift between the year 1999 and 2003 could be measured
since in 1999 a similar experiment has been done accidentally. Today the
frequency of the 1S--2S--transition is measured to 2466 061 413 187 127 Hz,
with an uncertainty of 18 Hz. The drift during the past 43 months is given
by 24 Hz, uncertainty about 50 Hz. This implies a change of -0.9 (2.9)
$10^{-15}$ per year.

Thus it is found that the prediction of about $2 \times 10^ {-14}$ per year
is presumably not realized. But further tests are going on.

Nevertheless we have to think what might be the reason that no change seems
to be there on the level of $10^{-14}$. Of course, there is the possibility
that the astrophysics result is wrong. Further tests to check this are being
prepared. But is could also be that a cancellation takes place. The time change
$ \left(\dot{\alpha}_s / \alpha_s \right)
$ receives 2 contributions, one by
$\left( \dot{\alpha} / \alpha \right)$, but also one by
$\left( \dot{\Lambda}_{{\rm GUT}} / \Lambda_{{\rm GUT}} \right)$.
If both are present, one
could have a suppression such that e. g.
$\left( \dot{\Lambda} / \Lambda \right)$ is not $30 \cdot
\left( \dot{\alpha} / \alpha \right)$, but only $3 \times \left(
\dot{\alpha} / \alpha \right)$. This would imply that in the experiment of
Haensch et al. the effect is there at the level of few
$\times 10^{-15}$ / year.

Tests to look for such an effect are being prepared. But it will take at
least one year, before results are known. It might also be that the
astrophysics observations are wrong. Recently new observations were
published, indicating a null-effect \cite{Quart}.

I like to thank S. Narison for arranging this splendid meeting in the
capital of Madagascar.

\end{document}